# Origin of layer number dependent linear and nonlinear optical properties of two-dimensional graphene-like SiC

You-Zhao Lan*[1]

*Institute of Physical Chemistry, College of Chemistry and Life Sciences, Zhejiang Normal University, Zhejiang, Jinhua 321004, China*

**Abstract**

On the basis of the sum-over-states calculations, we theoretically discuss the physical origin of the dielectric constants [$\varepsilon(\omega)$] and second harmonic generation coefficients [$\chi^{(2)}(\omega)$] of the *ABA*-stacked two-dimensional graphene-like silicon carbide (2D-SiC) with the number of layers up to 5. It is found that the intensities of the pronounced peaks of both $\varepsilon(\omega)$ and $\chi^{(2)}(\omega)$ exhibit a clear layer number dependence. For the light polarization *parallel* to the 2D-SiC plane, the monolayer SiC (ML-SiC) and multilayer SiC (MuL-SiC) have very similar pronounced peak positions of $\varepsilon(\omega)$, which are attributed to the $\pi \to \pi^*$ and $\sigma \to \sigma^*$ transitions. However, for the light polarization *perpendicular* to the 2D-SiC plane, a characteristic peak is found for the MuL-SiC at about 4.0 eV, except that the allowed $\pi \to \sigma^*$ and $\sigma \to \pi^*$ transition peaks are found for both ML-SiC and MuL-SiC in the high-energy region (> 8 eV). This characteristic peak is attributed to the *interlayer* $\pi \to \pi^*$ transition which does not exist for the ML-SiC, and at this peak position, the ML-SiC has a weak dark exciton based on the mBJ calculation within the Bethe-Salpeter equation framework. For $\chi^{(2)}(\omega)$, the single-particle transition channels based on the three-band terms dominate the second harmonic generation process of both ML-SiC and MuL-SiC and determine the size and sign of $\chi^{(2)}(\omega)$. In the ultraviolet visible region, the purely interband motion and intraband motion of electrons competitively determine the size and sign of $\chi^{(2)}(\omega)$. For the light polarization *perpendicular* to the 2D-SiC plane, the intraband motion of electrons modulated more dramatically the interband motion than for that *parallel* to the 2D-SiC plane. For ML-SiC, our results with and without including excitonic effects agree well with the previous ones based on the real time first principle calculations and the local density approximation calculations.

---

[1] Corresponding author: Youzhao Lan; Postal address: *Institute of Physical Chemistry* , College of Chemistry and Life Sciences, Zhejiang Normal University, Zhejiang, Jinhua 321004, China; Fax: +086 579 82282269; E-mail address: lyzhao@zjnu.cn

1. **Introduction**

The two-dimensional graphene-like silicon carbide (2D-SiC) has attracted much attention owing to its different electronic properties from graphene. It is well known that the monolayer graphene exhibits a zero bandgap and has unusual two-dimensional Dirac-like electronic excitations [1]. However, the monolayer SiC (ML-SiC) has a large direct bandgap of 2.54 eV based on the density functional theory (DFT) calculations [2–4] within both the local density approximation (LDA) and the generalized gradient approximation (GGA). Within the many-body *GW* approximation [5], the bandgap of ML-SiC will increase to about 4.0 eV (Ref.[6–10]). The ML-SiC would be suitable for applications in optoelectric devices such as light emitting diodes and solar cells owing to its having a direct bandgap. Obviously, isolating the ML-SiC is difficult in practical experiments, and up to now, only ultrathin two-dimensional SiC flakes with a thickness of 0.5−1.5 nm have been obtained [11]. Thus, it is worthwhile to theoretically characterize the multilayer SiC (MuL-SiC).

It has been shown that the multilayer (or few-layer) graphene exhibit the notable electronic and optical properties. The bilayer graphene has a tunable bandgap in the range of 70-250 meV with the external electric field applied perpendicular to the bilayer plane [12,13]. In infrared conductivity measurements [14], the principal transitions exhibit a systematic energy-scaling behavior with the number of layers (*N*) increasing from 1 to 8. The trilayer graphene with *ABA*- and *ABC*-stacking possesses the stacking-order dependent transport properties [15]. The interlayer oriented misalignment [16] leads to dramatically different electronic and optical properties, such as energy-renormalization of the low-energy spectra [17] and twist-angle dependent spectra [18], from the monolayer graphene owing to the complicated interaction between the misaligned layers. Undoubtedly, similar effects of the oriented misalignment [16–18], stacking-order [15,19,20], and *N*-dependence [14,21–23] on the electronic and optical properties will occur for the MuL-SiC. The *ABA*-stacked MuL-SiC transforms to the indirect bandgap semiconductor owing to the weak interaction between layers, and the corresponding bandgap is less than that of ML-SiC [3] and gradually decreases with an increase of *N*. Interestingly, although the *ABA*-stacking [3] leads to a direct-to-indirect bandgap transition from the ML-SiC to the MuL-SiC, the oriented misalignment will induce an indirect-to-direct bandgap transition in the bilayer/trilayer 2D-SiC [24]. Overall, with an increase of *N* of 2D-SiC, the electronic and optical properties become more and more complicated but intriguing.

In this work, we report the linear and nonlinear optical properties of the graphene-like 2D-SiC with *N* up to 5. We present the *N*-dependence of the dielectric constants [$\varepsilon(\omega)$] and second harmonic generation coefficients [$\chi^{(2)}(\omega)$] for the light polarizations parallel and perpendicular to the plane. The electronic origin of absorption peaks in $\varepsilon(\omega)$ is carefully analyzed by extracting the optical matrix elements, and the $\chi^{(2)}(\omega)$ is understood by three ways, namely, resonant enhanced contributions, single particle channel contributions, and contributions from interband and intraband motions of electrons. The paper is organized as follows. In Sec. 2, we present the computational details including the geometry optimizations and calculations of band structure and optical properties. In Sec. 3, we first make a comparison between the PBE with scissor correction and mBJ-PW, then discuss the physical origin of $\varepsilon(\omega)$ and $\chi^{(2)}(\omega)$, and finally compare our results with and without including the excitonic effects with the previous theoretical ones for ML-SiC. In Sec. 4, the conclusions are given.

## 2. Computational details and formulism

### 2.1 Geometry

The crystal structure of two-dimensional layered SiC is shown in Fig. 1. For the ML-SiC, theoretical researches on its geometry and electronic properties [3,6,8,11,25–27] have shown that ML-SiC has a planar structure with the $D_{3h}$ symmetry and is a semiconductor with a large gap of about 2.54 eV. For the MuL-SiC, we show interest in the *ABA*-stacking which has two kinds of stacking configurations; one is a carbon atom opposite a silicon atom, the other is a carbon (silicon) atom opposite a carbon (silicon) atom. According to Kaplan *et al.*[27,28], the former configuration is more favorable in energy than the latter one; thus in our present work we only focus on the former configuration (*i.e.*, a carbon atom opposite a silicon atom). We optimized the structures of ML- and MuL-SiC by using the density functional theory approach within the GGA of the Perdew, Burke, and Ernzerhof (PBE) functional [29,30], as implemented in the CASTEP code [31]. A *k*-point mesh of 10×10×1, a force threshold of 0.01 eV/Å, and a stress threshold of 0.02 GPa were used for optimizations. The optimized distance between the adjacent layers is 3.73 Å, in consistent with the experimental determination [11], and thus the largest thickness of our *ABA*-stacking is about 1.5 nm (*i.e.*, 5 layers) which lies in the thickness range of ultrathin flakes [11]. A vacuum spacing of 15 Å was used to assure a negligible interaction between the slabs. With a relaxation of the unit cell, the optimized lattice lengths ($a = b$) are 3.10 Å (Fig.1), in agreement with previous theoretical researches [3,6].

*2.2 Band structure*

Our present calculations were performed by solving the self-consistent Kohn-Sham equations within the GGA. The highly accurate all-electron full potential linearized augmented plane wave (FP-LAPW) method [32,33] implemented in ELK code [34] was used. Since the conventional LDA and GGA functionals generally underestimate the band gap of solids, we used the mBJ exchange potential [35] combined with the LDA based Perdew and Wang (PW) correlation potential [36]. The use of the mBJ-PW functional gives a good agreement between theoretical and experimental band gaps for a majority of solids [35]. Note that in the use of mBJ-PW functional, there is an important tunable parameter $c$ which is defined as [35]:

$$c = \alpha + \beta \left( \frac{1}{V_{cell}} \int_{cell} \frac{|\nabla \rho(r)|}{\rho(r)} d^3r \right)^{1/2} \qquad (1)$$

where $\alpha$ and $\beta$ are two free parameters, $V_{cell}$ is the unit cell volume, and $\rho$ is the electron density. For $c = 1$, we will go back to the original BJ potential [37]. By comparing with both the experimental band gaps and the theoretical *GW* ones, Tran and Blaha have shown that the optimal $c$ value lies in the range of 1.1 – 1.3 and 1.4 – 1.7 for solids with small ( < 1.0 eV) and large ( > 1.0 eV) bandgaps, respectively. Herein, we decided an optimal $c$ value on the basis of the *GW* band gaps. Table 1 lists the bandgaps of ML-SiC based on different GGA and *GW* calculations. For ML-SiC, the PBE bandgap is 2.55 eV while the *GW* bandgap ranges from 3.80 to 4.40 eV depending on different *GW* approximations. On the basis of the *GW* bandgaps, we obtain an optimal $c$ value of 1.5 which leads to an mBJ-PW bandgap of 3.93 eV for ML-SiC. This optimal $c$ value was used to calculate the bandgaps of MuL-SiC (*i.e.*, 2L-, 3L-, 4L-, and 5L-SiC) because MuL-SiC has a similar band structure to ML-SiC based on the first principle calculation [3]. The band structure calculation will be followed by the optical calculation which requires a dense $k$-point mesh of 60×60×1 (section 2.3), so this dense mesh is also used for the band structure calculation to assure the converged energy.

*2.3 Optical properties*

We calculated the linear and nonlinear optical properties within the independent particle approximation (IPA) [38,39]. The expressions for the linear optical properties are simple and well known. For the nonlinear optical properties, we show interest in the second harmonic generation (SHG). The static-divergence-free expressions for SHG are much more complicated than those for the linear optical response. The expressions for SHG were originally developed by Sipe and Ghahramani [39] and widely used to calculate the SHG of semiconductors [40–45]. All the expressions for the linear and SHG properties are as follows (Eqs. 2–6):

For linear optical response:

$$\chi_{\text{inter}}^{ba}(-\omega,\omega) = \frac{1}{\Omega}\sum_{nmk}\frac{r_{mn}^{b}r_{nm}^{a}f_{mn}}{\omega_{nm}-\omega} \qquad (2)$$

$$\varepsilon_{ba}(\omega) = \delta_{ba} + 4\pi\chi_{\text{inter}}^{ba}(-\omega,\omega) = \varepsilon_1(\omega) + i\varepsilon_2(\omega) \qquad (3)$$

where $\omega_{mn} = \omega_m - \omega_n$ is the energy difference between the $m^{\text{th}}$ and $n^{\text{th}}$ bands, $f_{mn} = f_m - f_n$ is the difference of the Fermi distribution functions, the indices of $a$ and $b$ are Cartesian directions, the band indices $m$, $n$ are different because $r_{mn}$ [= $p_{mn}/(im\omega_{mn})$] is defined to be zero unless $n \neq m$ (Ref.[38]), and $\Omega$ is the cell volume. The $k$-dependence of $r_{mn}$ is suppressed for clarity. The 'inter' indicates the interband contribution.

For SHG:

$$\chi_{\text{inter}}^{cba}(-2\omega,\omega,\omega) = \frac{e^3}{\hbar^2\Omega}\sum_{mnlk}\frac{r_{mn}^c\{r_{nl}^b r_{lm}^a\}}{(\omega_{lm}-\omega_{nl})}\left[\frac{2f_{mn}}{\omega_{nm}-2\omega}+\frac{f_{nl}}{\omega_{nl}-\omega}+\frac{f_{lm}}{\omega_{lm}-\omega}\right]$$

$$=\frac{e^3}{\hbar^2\Omega}\sum_{mnlk}\frac{r_{mn}^c\{r_{nl}^b r_{lm}^a\}}{(\omega_{lm}-\omega_{nl})}\frac{2f_{mn}}{\omega_{nm}-2\omega}+\frac{e^3}{\hbar^2\Omega}\sum_{nmlk}\frac{r_{mn}^c\{r_{nl}^b r_{lm}^a\}}{(\omega_{lm}-\omega_{nl})}\frac{f_{nl}}{\omega_{nl}-\omega} \quad (4)$$

$$+\frac{e^3}{\hbar^2\Omega}\sum_{nmlk}\frac{r_{mn}^c\{r_{nl}^b r_{lm}^a\}}{(\omega_{lm}-\omega_{nl})}\frac{f_{lm}}{\omega_{lm}-\omega}$$

$$\chi_{\text{intra}}^{cba}(-2\omega,\omega,\omega) = \frac{e^3}{\hbar^2\Omega}\sum_{mnlk}\omega_{nm}r_{mn}^c\{r_{nl}^b r_{lm}^a\}\left\{\frac{f_{ml}}{\omega_{lm}^2(\omega_{lm}-\omega)}-\frac{f_{\ln}}{\omega_{nl}^2(\omega_{nl}-\omega)}\right\}$$

$$+2\frac{e^3}{\hbar^2\Omega}\sum_{mnlk}\frac{f_{mn}r_{mn}^c\{r_{nl}^b r_{lm}^a\}(\omega_{nl}-\omega_{lm})}{\omega_{nm}^2(\omega_{nm}-2\omega)}-8i\frac{e^3}{\hbar^2\Omega}\sum_{mnk}\frac{f_{mn}r_{mn}^c\{\Delta_{nm}^b r_{nm}^a\}}{\omega_{nm}^2(\omega_{nm}-2\omega)} \quad (5)$$

$$\chi_{\text{mod}}^{cba}(-2\omega,\omega,\omega) = \frac{1}{2}\frac{e^3}{\hbar^2\Omega}\sum_{mnlk}\frac{f_{mn}}{\omega_{nm}^2(\omega_{nm}-\omega)}\left\{\omega_{ml}r_{\ln}^c\{r_{nm}^b r_{ml}^a\}-\omega_{\ln}r_{ml}^c\{r_{\ln}^b r_{nm}^a\}\right\}$$

$$+\frac{1}{2}i\frac{e^3}{\hbar^2\Omega}\sum_{mnk}\frac{f_{mn}r_{mn}^c\{r_{nm}^b\Delta_{nm}^a\}}{\omega_{nm}^2(\omega_{nm}-\omega)} \quad (6)$$

where the {} in $\{r_{nl}^b r_{lm}^a\}$ is defined to be $1/2(r_{nl}^b r_{lm}^a + r_{nl}^a r_{lm}^b)$ to make the expressions satisfy the intrinsic permutation symmetry and $\triangle_{nm}^a \equiv \upsilon_{nn}^a - \upsilon_{mm}^a$. The 'inter' also indicates the purely interband contribution. Both 'intra' and 'mod' indicate the contributions from the combined intra- and interband motion, and the 'intra' term describes the modulation of polarization by the intraband motion of electrons while the 'mod' term indicates the intraband motion modified by the polarization energy associated with the interband motion of electron. [41,44]

Calculations of the optical properties based on the expressions above require the band structure of system and the momentum matrix elements. The highly accurate all-electron FP-LAPW method was used to calculate the band structure and the corresponding optical matrix elements which are required by the sum-over-states (SOS) calculations of the optical properties. Since the SOS calculation of the linear and nonlinear optical properties depends on the number of empty states and the *k*-point mesh, as an example, we show in Fig.2 the convergence tests on the $\varepsilon_{1-xx}(\omega)$ and $|\chi^{(2)}_{xxx}(\omega)|$ of ML-SiC. For the *k*-point mesh, both the $\varepsilon_{1-xx}(\omega)$ and $|\chi^{(2)}_{xxx}(\omega)|$ converge with the 60×60×1 mesh, and for the number of empty states included in the SOS calculation, good convergences are also obtained for both the $\varepsilon_1(\omega)$ and $|\chi^{(2)}_{xxx}(\omega)|$. Therefore, for MuL-SiC, we adopted the 60×60×1 *k*-point mesh to calculate their band structures and the corresponding momentum matrix elements. Note that the number of empty states is set to 10 empty states *per* atom, for example, 40 empty states are used for 2L-SiC (four atoms in unit cell).

### 3. Results and discussion

*3.1 PBE with scissor correction versus mBJ-PW*

In previous calculations of optical properties of solids [41,42,44,46,47], the bandgap was often underestimated when compared to the experimental bandgap. There are two ways of disposing this underestimation, one is introducing a so-called scissor correction [47], the other is improving the band structure by developing new functionals [35,37]. In the former, the bandgap is shifted by a factor of $\Delta\omega$ that is generally selected as the difference between theoretical and experimental gaps and the momentum matrix elements are corrected by $p_{mn} = p_{mn}(1+f_{mn}\Delta\omega/\omega_{nm})$. In this section, we make comparisons among the PBE without scissor correction (PBEwosc), the PBE with scissor correction (PBEwsc), and the mBJ-PW calculations. As an example, figure 3 shows the $\varepsilon_{1-xx}(\omega)$ and $|\chi^{(2)}_{xxx}(\omega)|$ of ML-SiC based on the PBEwosc, PBEwsc, and mBJ-PW with the all-electron FP-LAPW calculations. According to the PBE and mBJ-PW bandgaps (table 1), we used a scissor correction of 1.37 eV for ML-SiC. We can see firstly from Fig.3 that for both $\varepsilon_{1-xx}(\omega)$ and $|\chi^{(2)}_{xxx}(\omega)|$ the PBEwsc spectra have an obvious shift of peak positions when compared to the PBEwosc ones. The shift size is close to the scissor value. This is due to the fact that the scissor correction ($\Delta\omega$) is a rigid shift [48] for $\omega_{nm}$, and then leads to a shift of peak positions from $\omega$ to $\omega+\Delta\omega$.

Secondly, the mBJ-PW spectra are very similar to the PBEwsc ones in terms of the line shape, peak positions, and even the size of $\varepsilon_{1-xx}(\omega)$ and $|\chi^{(2)}_{xxx}(\omega)|$. As a possible interpretation for this similarity, we show in Fig. 3c the calculated PBE and mBJ-PW band structures based on the FP-LAPW method. The mBJ-PW bands compared to the PBE ones mainly show a rigid shift of energies (*i.e.*, up for conduction bands and down for valence bands) and the *k*-dispersion of energy bands based on these two methods are very similar. For MuL-SiC, we obtain similar results (Fig. S1). In the following sections, as an application of new functional (mBJ-PW) to the calculation of nonlinear optical property, the mBJ-PW results are used for discussions and the PBE results are only shown for comparisons.

*3.2 Linear optical response*

For the linear optical properties of ML-SiC and MuL-SiC, we show interest in their dielectric constants because these systems belong to the semiconductor with a large bandgap (table 1). Figure 4 shows the real and imaginary parts of the mBJ dielectric constants with the polarization along two directions [*i.e.*, parallel ($\varepsilon_{xx}$) and perpendicular ($\varepsilon_{zz}$) to the 2D-SiC plane] for ML-SiC and MuL-SiC, and for clarity the static mBJ dielectric constants [$\varepsilon_1(0)$] of ML-SiC and MuL-SiC are given in table 2. For ML-SiC, the PBE results are very close to previous LDA ones [4]. As shown in Fig.4, the dispersions of $\varepsilon_1(\omega)$ and $\varepsilon_2(\omega)$ exhibit two pronounced features. One is a clear layer-dependence of the size of both $\varepsilon_{xx}$ and $\varepsilon_{zz}$. As shown in table 2, the $\varepsilon_1(0)$ values clearly increase with an increase of *N*. To understand this variation trend, we turn to the Eqs.2 and 3 which indicates that $\varepsilon_1(0)$ is inversely and directly proportional to the transition energy ($\omega_{nm}$) between the valance and conductor bands and to the norm of position matrix element ($r_{mn}$), respectively, [*i.e.*, $\varepsilon_1(0) \propto f_{mn}|r_{mn}|^2/(\Omega\omega_{nm})$] (Ref.[39]). For example, the $\omega_{nm}$ and $|r_{mn}|^2/\omega_{nm}$ (*xx* and *zz* components) for the allowed transitions between the frontier states at the K *k*-point of the Brillouin zone are given in table 3, while the final contribution to $\varepsilon_1(0)$ [*i.e.*, $\sum_{mn}|r_{mn}|_{xx}^2/(\Omega\omega_{nm})$] is listed in table 2 for convenience of viewing the *N*-dependence. For ML-SiC with 8 valence electrons, we list the transition from the 4$^{th}$ (the highest occupied state) to the 5$^{th}$ (the lowest unoccupied state) state. For 2L-SiC with 16 valence electrons, the transitions from the 7$^{th}$ and 8$^{th}$ (the highest occupied state) to the 9$^{th}$ (the lowest unoccupied state) and 10$^{th}$ are given. Two occupied states (7$^{th}$ and 8$^{th}$) are considered for 2L-SiC because they evolve from the 4$^{th}$ state of the stacked ML-SiC, which can be seen from Fig.5 that displays the band structures and partial density of states (PDOS) of ML-SiC and 2L-SiC. Similar cases are for 3L-SiC, 4L-SiC, and 5L-SiC. As shown in tables 1 and 3, both the bandgaps and the *allowed* transition energies decrease with an increase of *N*. The variation trend of final contributions [*i.e.*, $\sum_{mn}|r_{mn}|^2/(\Omega\omega_{nm})$] (table 2) is consistent with that of $\varepsilon_1(0)$ even if the cell volume ($\Omega$) increases with an increase of *N*. Note that the used $\Omega$ value will dramatically influence the size of $\varepsilon(\omega)$ but not the peak positions and envelopes in the dispersion of dielectric constants [4], and thus we used the same vacuum spacing of 15 Å for ML-SiC and MuL-SiC to eliminate the effect due to different vacuum spacings on the layer dependence of $\varepsilon$.

The other feature is that, for $\varepsilon_{xx}$, both ML-SiC and MuL-SiC have two characteristic peaks at about 4.0 and 9.0 eV, while for $\varepsilon_{zz}$, MuL-SiC has a peak at about 4.0 eV which is not shown for ML-SiC. For the $\varepsilon_{xx}$ of ML-SiC, in agreement with the previous report [4], two characteristic peaks (4.0 and 9.0 eV) can be attributed to the $\pi\rightarrow\pi^*$ and $\sigma\rightarrow\sigma^*$ transitions, respectively, which can be seen from the PDOS (Fig. 5). Similar results are obtained for the $\varepsilon_{xx}$ of MuL-SiC. For the $\varepsilon_{zz}$ of ML-SiC, the $\pi\rightarrow\pi^*$ transition is forbidden, which can be seen from the $r_{mn}$ value (table 3), while only $\pi\rightarrow\sigma^*$ and $\sigma\rightarrow\pi^*$ transitions are allowed, and thus the absorption peaks only appear in the high-energy region (> 8 eV). Interestingly, for the $\varepsilon_{zz}$ of MuL-SiC, the $\pi\rightarrow\pi^*$ transitions are allowed, which leads to the presence of absorption peaks at about 4.0 eV. These allowed $\pi\rightarrow\pi^*$ transitions arise from the weak interlayer interaction. As shown in Fig. 5, the band structure of 2L-SiC exhibit a slight split when compared to that of ML-SiC. This split is ascribed to the weak interlayer interaction; otherwise, the corresponding bands should be degenerate. Furthermore, according to the $|r_{mn}|_{zz}^2/\omega_{nm}$ values (tables 3 and S1), the band structures (Fig.S1), and the PDOS (Figs.5), the $\pi\rightarrow\pi^*$ transitions of MuL-SiC should be viewed as the *interlayer* $\pi\rightarrow\pi^*$ transition relative to the *intralayer* $\pi\rightarrow\pi^*$ transition of ML-SiC. For example, for 2L-SiC transitions among the frontier states (7[th], 8[th], 9[th], and 10[th]) show that only the 7[th] $\rightarrow$ 10[th] transition has a nonzero $|r_{mn}|_{zz}^2/\omega_{nm}$ value (table 3). The 7[th] and 10[th] states mainly come from the atoms of layers A and B, respectively, as shown in the PDOS of 2L-SiC (Fig.5). A similar behavior exist in the $\varepsilon_{2\text{-}zz}(\omega)$ of graphene [4] and graphite [49].

*3.3 Second order optical response*

Figure 6 shows the $|\chi^{(2)}(\omega)|$ (a.u.) along two polarization directions (*x* and *z*) for ML-SiC and MuL-SiC. Note that for ML-SiC, 3L-SiC, and 5L-SiC with a $D_{3h}$ symmetry, only 4 elements for the second order susceptibility tensor (*i.e.*, *xxx*, *zxx*, *xxz*, and *xzx*) are nonzero and they obey the equality of *xxx* = –*zxx* = –*xxz* = –*xzx*, while for 2L-SiC and 4L-SiC with a $C_{3v}$ symmetry which have 11 nonzero elements, we focus on two elements (*i.e.*, *xxx* and *zzz*). As shown in Fig.6, both $|\chi^{(2)}_{xxx}(\omega)|$ and $|\chi^{(2)}_{zzz}(\omega)|$, similar to $\varepsilon(\omega)$ (Fig.4), exhibit a clear layer dependence, that is, $|\chi^{(2)}(\omega)|$ increases with an increase of *N*. Obviously, variation trends of either bandgaps or transition energies will be one of main factors resulting in this *N*-dependence. However, since expressions (Eqs. 4–6) for $\chi^{(2)}(\omega)$ are more complicated than that (Eq.2) for $\chi^{(1)}(\omega)$, we will understand the calculated $\chi^{(2)}(\omega)$ in the following three ways.

Firstly, for $|\chi^{(2)}_{xxx}(\omega)|$, we observe a single peak at about 2.2 eV and multiple peaks at about 4.5 eV, while for $|\chi^{(2)}_{zzz}(\omega)|$, the distribution of peaks is wider than that for $|\chi^{(2)}_{xxx}(\omega)|$. These peaks can be attributed to the one- and/or two-photon resonances according to Eqs. 4–6. For example, in Eq. 4, the first summation term contributes to the two-photon resonance due to $\omega_{nm}-2\omega$, while the second and third terms belong to the one-photon resonance due to $\omega_{nl}-\omega$ and $\omega_{lm}-\omega$, respectively. To determine the type of resonances for these peaks, we shall consult the dielectric function which depends on the $\omega_{mn}-\omega$ term [7,44,45,50–52]. As shown in Fig. 4, for $\varepsilon_{2\text{-}xx}$, the absorption peaks mainly locate at about 4.4 and 9.0 eV. So, for $|\chi^{(2)}_{xxx}(\omega)|$, peaks at about 2.2 eV arise from the two-photon resonance, while those at about 4.5 eV arise from the one- and/or two-photon resonances.

Secondly, $\chi^{(2)}(\omega)$ of Eqs. 4–6 can be decomposed into two- and three-band terms based on the number of bands included in summation [39,53]. For example, all three summations in Eq.4 are considered as three-band term due to a summation on *m*, *n*, and *l* bands, while the third term of Eq. 5 is the two-band term due to a summation on *m* and *n* bands. Figures 7b1, 7c1, 7b2, and 7c2 display the real and imaginary parts of $\chi^{(2)}_{xxx}(\omega)$ coming from two- and three-band terms. Note that the two-band contribution is magnified by $\times 10^3$ in Figs. 7b1 and 7b2 for convenience of comparison. Clearly, both two- and three-band contributions exhibit the $N$-dependence. Comparing Figs. 7(b1, b2, c1, and c2) to Figs. 7(a1 and a2), we can see that the three-band terms have a much larger contributions to $\chi^{(2)}_{xxx}(\omega)$ than the two-band terms and determine the sign and size of real and imaginary parts of $\chi^{(2)}_{xxx}(\omega)$. With the restriction of the difference of Fermi factors (*i.e.*, $f_{mn}$), three-band terms describe the virtual-electron transitions from one valence band to two conduction bands, or the virtual-hole transitions from one conduction band to two valence bands [50,51,53]. For example, the first summation term of Eq.4 is described by the virtual-electron transition when $m$ is valence band and both $n$ and $l$ are conduction bands. Two-band terms (the third term of Eq.5 and the second term of Eq.6) describe an interband transition from valence band $m$ to conduction band $n$ and two intraband motions ($\triangle^a_{mn}$) in bands $m$ and $n$. As shown in Figs.7 (b1, b2, c1, and c2), three-band contributions are almost $10^3$ times as large as two-band ones; therefore it is suggested that the single-particle processes corresponding to the three-band terms would mainly contribute to the second harmonic generation process of ML-SiC and MuL-SiC.

Thirdly, we understand the $\chi^{(2)}(\omega)$ values in terms of the decomposition of [$\chi^{(2)}_{inter}(\omega)$, $\chi^{(2)}_{intra}(\omega)$, and $\chi^{(2)}_{mod}(\omega)$] (Eqs.4–6). The real and imaginary parts of $\chi^{(2)}_{inter}(\omega)$, $\chi^{(2)}_{intra}(\omega)$, and $\chi^{(2)}_{mod}(\omega)$ are shown in Fig.8 for $\chi^{(2)}_{xxx}(\omega)$. For the real part of $\chi^{(2)}_{xxx}(\omega)$, we may divide the spectra into three regions, namely, $\hbar\omega < 2.0$ eV, $2.0$ eV $< \hbar\omega < 4.0$ eV, and $\hbar\omega > 4.0$ eV. For $\hbar\omega < 2.0$ eV, we can see that $\chi^{(2)}_{intra}(\omega)$ determines the size and sign of $\chi^{(2)}_{xxx}(\omega)$ but $\chi^{(2)}_{inter}(\omega)$ has a larger and larger contribution to $\chi^{(2)}_{xxx}(\omega)$ (see peak height at 2.0 eV) as an increase of $N$. For $2.0$ eV $< \hbar\omega < 4.0$ eV, $\chi^{(2)}_{intra}(\omega)$ also determines the size and sign of $\chi^{(2)}_{xxx}(\omega)$ because $\chi^{(2)}_{inter}(\omega)$ and $\chi^{(2)}_{mod}(\omega)$ almost cancel out in summations. For $\hbar\omega > 4.0$ eV, however, $\chi^{(2)}_{inter}(\omega)$ determine the size and sign of $\chi^{(2)}_{xxx}(\omega)$ because $\chi^{(2)}_{intra}(\omega)$ and $\chi^{(2)}_{mod}(\omega)$ almost cancel out in summations.

For the imaginary part of $\chi^{(2)}_{xxx}(\omega)$, we may divide the spectra into two regions by using 3.0 eV as a split point. For $\hbar\omega < 3.0$ eV, $\chi^{(2)}_{intra}(\omega)$ determines the size and sign of $\chi^{(2)}_{xxx}(\omega)$ but $\chi^{(2)}_{inter}(\omega)$ has a larger and larger contribution to $\chi^{(2)}_{xxx}(\omega)$ (see also peak height at 2.0 eV) as an increase of *N*. For $\hbar\omega > 3.0$ eV, $\chi^{(2)}_{inter}(\omega)$ and $\chi^{(2)}_{intra}$ together determine the size, sign, and peak positions of $\chi^{(2)}(\omega)$. $\chi^{(2)}_{inter}(\omega)$ and $\chi^{(2)}_{intra}$ lead to two adjacent peaks at about 4.0 eV (*e.g.*, see arrows in 2L-SiC of Fig.8). Note that $\chi^{(2)}_{mod}(\omega)$ only has a few contributions at around 4.0 eV. According to the Eq.6, $\chi^{(2)}_{mod}(\omega)$ consists of one three-band term and one two-band term. Since this two-band term only has very little contribution to $\chi^{(2)}_{xxx}(\omega)$ as mentioned above, $\chi^{(2)}_{mod}(\omega)$ should mainly come from the three-band term and the corresponding peak is due to the one-photon resonance based on the $\omega_{nm}-\omega$ factor of Eq.6. Therefore, in the whole ultraviolet visible region (1.5 eV – 6.0 eV), $\chi^{(2)}_{inter}(\omega)$ and $\chi^{(2)}_{intra}(\omega)$ competitively determine the size and sign of $\chi^{(2)}_{xxx}(\omega)$. The $\chi^{(2)}_{intra}(\omega)$ has more significant contributions to $\chi^{(2)}(\omega)$ than the $\chi^{(2)}_{inter}(\omega)$ overall, which means that the intraband motion of electrons dramatically modulate the interband motion of electrons. Finally, for $\chi^{(2)}_{zzz}(\omega)$, although only 2L-SiC and 4L-SiC have nonzero values due to the limit of symmetry, all these three understandings for $\chi^{(2)}_{xxx}(\omega)$ are appropriate for $\chi^{(2)}_{zzz}(\omega)$ (see Figs. S2 and S3).

*3.4 Comparison with previous theoretical results for ML-SiC.*

There are some theoretical results for the $\varepsilon(\omega)$ and $\chi^{(2)}(\omega)$ of ML-SiC based on the real-time (RT) first-principle calculation [7] and the Kohn-Sham density functional theory (DFT) [4,9,52]. The excitonic effect on the $\varepsilon(\omega)$ and $\chi^{(2)}(\omega)$ of ML-SiC has also been investigated [7,9]. Figures 9 and 10 separately display the $\varepsilon_{2-xx/zz}(\omega)$ and $\chi^{(2)}_{xxx}(\omega)$ of ML-SiC based on different theoretical calculations including RT calculations [7] within IPA and GW + Bethe-Salpeter equation (GWBSE) framework, both mBJ and PBEwsc within BSE framework, as well as mBJ, PBEwsc, and PBEwosc within IPA. Note that we only compare the line shape and the *relative* height and position of peaks of spectra because for ML-SiC the *arbitrary* vacuum spacing assuring no interlayer interaction can be used to calculate the cell volume ($\Omega$) in d$k$ integration. Using the same effective thickness (about 3.75 Å) in our PBEwosc-IPA calculations, we can obtain similar intensity of peaks to the LDA-IPA ones without the scissor correction [4,52] (not shown in Fig.9). As shown in Fig. 9a, within the BSE framework, for $\varepsilon_{2\text{-}xx}(\omega)$, our mBJ and PBEwsc results are in good agreement with the RT-GWBSE ones [7] in terms of the line shape and the relative height and position of peaks. Differences in the absolute position of peaks would be ascribed to the used different strategies (*i.e.*, mBJ, PBEwsc, RT-GW) followed by the BSE calculation. Within the IPA (Fig. 9b), our PBEwosc results are in good agreement with the RT ones. There is an interesting cancellation in energy shifts caused by GW and BSE [7], as a result, the GW-BSE spectrum have very similar positions of main peaks to the IPA one for both RT and our DFT calculations. This cancellation is also observed for the mBJ-BSE spectrum.

For the $\varepsilon_{2\text{-}zz}(\omega)$ of ML-SiC, no exciton corrected spectrum has been reported. As shown in Figs. 9(c and d), in addition to similar features to $\varepsilon_{2\text{-}xx}(\omega)$, we observe a peak at about 4.6 eV in the mBJ-BSE spectrum but not in the PBEwsc-BSE spectrum. This peak can be attributed to a dark exciton because no optically allowed transition is observed at around 4.0 eV in the $\varepsilon_{2\text{-}zz}(\omega)$ of ML-SiC (table 3 and Fig. 4d). For $\chi^{(2)}_{xxx}(\omega)$, we compare our results with the RT and LDA ones [7,52]. As shown in Fig. 10, our mBJ spectra are close to the RT and LDA ones in terms of the relative position of peaks. Differences in the line shape and intensities of main peaks should be due to the inclusion of the intraband contribution in our calculations. Similarly, since there is a cancellation in energy shift between GW and BSE mentioned by Attaccalite *et al.* [7], our PBEwosc-IPA spectrum (Fig. 10c) is similar to both the RT-GWBSE and RT-IPA spectra.

## 4. Conclusions

We have presented the *N*-dependence and physical origin of $\varepsilon(\omega)$ and $\chi^{(2)}(\omega)$ of the *ABA*-stacked 2D-SiC with *N* up to 5 on the basis of the first principle SOS calculations. The mBJ exchange potential approximation can efficiently reproduce the *GW* bandgap for 2D-SiC and has a similar performance to the scissor correction for calculating the optical properties. For the light polarization perpendicular to the 2D-SiC plane, the $\pi \rightarrow \pi^*$ transition is only allowed between layers, and thus it does not occur for the ML-SiC. We might distinguish the ML-SiC and MuL-SiC by using the $\varepsilon_{2\text{-}zz}(\omega)$ because the MuL-SiC has a characteristic peak at about 4.0 eV which is not shown in the $\varepsilon_{2\text{-}zz}(\omega)$ of ML-SiC. The interband and intraband motions of electrons modulate each other. The intraband motion of electrons has a significant influence on the interband motion and the modulated interband contributions [$\chi^{(2)}_{\text{intra}}(\omega)$] determine the size and sign of the $\chi^{(2)}(\omega)$ overall. Finally, our present results are based on the *ABA*-stacking, however, other elements, such as the stacking order and oriented misalignment, will definitely influence the *N*-dependence and physical origin of the optical properties and thus more efforts are needed in the future.

**Acknowledgements**

We appreciate the financial supports from Natural Science Foundation of China Project 21303164 and the computational supports from Zhejiang Key Laboratory for Reactive Chemistry on Solid Surfaces. The numerical calculations in this paper have been done on the supercomputing system in the Supercomputing Center of University of Science and Technology of China.

Table 1. Bandgaps (eV) of ML-SiC and MuL-SiC based on different theoretical calculations.

| Methods | ML-SiC | 2L-SiC | 3L-SiC | 4L-SiC | 5L-SiC |
|---|---|---|---|---|---|
| PBE | 2.57 | 2.20 | 2.23 | 2.06 | 1.98 |
| mBJ-PW | 3.93 | 3.47 | 3.42 | 3.29 | 3.22 |
| $GW$ [a] | 4.19 [b], 4.42 [c], 3.88 [d], 3.96 [e], 3.90 [f] | | | | |

[a] for ML-SiC only; [b] Reference [6] $GW^0$ (12×12×1 $k$-point mesh); [c] Reference [9] $G^0W^0$ (18×18×1 $k$-point mesh); [d] Reference [10] $G^0W^0$; [e] Reference [7] $G^0W^0$; [f] Reference [8] $GW^0$

Table 2. Static dielectric constants [$\varepsilon_1(0)$ defined in Eq.3] of ML-SiC and MuL-SiC.

| | ML-SiC | 2L-SiC | 3L-SiC | 4L-SiC | 5L-SiC |
|---|---|---|---|---|---|
| $\varepsilon_{1\text{-}xx}(0) = \varepsilon_{1\text{-}yy}(0)$ | 2.00 | 2.64 | 3.07 | 3.38 | 3.61 |
| $\varepsilon_{1\text{-}zz}(0)$ | 1.46 | 1.92 | 2.24 | 2.46 | 2.63 |
| $\varepsilon_1(0)$ [a] | 1.82 | 2.40 | 2.79 | 3.07 | 3.28 |
| $\Omega$ (a.u.) | 842.59 | 1051.31 | 1261.09 | 1470.52 | 1679.96 |
| $\sum_{mn}|r_{mn}|_{xx}^2/(\Omega\omega_{nm})$ ($\times 10^{-2}$ a.u.) [b] | 3.51 | 5.89 | 7.49 | 8.65 | 9.55 |
| $\sum_{mn}|r_{mn}|_{zz}^2/(\Omega\omega_{nm})$ ($\times 10^{-3}$ a.u.) | 0.00 | 4.21 | 7.25 | 8.90 | 10.02 |

[a] $\varepsilon_1(0) = 1/3[2\varepsilon_{1\text{-}xx}(0) + \varepsilon_{1\text{-}zz}(0)]$

[b] Contributions from the K $k$-point of the Brillouin zone (Fig.1)

Table 3. $\omega_{nm}$ (a.u.) and $|r_{nm}|^2/\omega_{nm}$ (a.u.) for the transitions between the frontier states at the K $k$-point (Fig. 5). For clarity, only allowed transitions are listed. All the "0.00"s indicate the value is less than $10^{-5}$. Similar results for 4L-SiC and 5L-SiC are given in table S1.

| m→n | $\omega_{nm}$ | $|r_{mn}|_{xx}^2/\omega_{nm}$ | $|r_{mn}|_{zz}^2/\omega_{nm}$ | m→n | $\omega_{nm}$ | $|r_{mn}|_{xx}^2/\omega_{nm}$ | $|r_{mn}|_{zz}^2/\omega_{nm}$ |
|---|---|---|---|---|---|---|---|
| ML-SiC | | | | 3L-SiC | | | |
| 4→5 | 0.1504 | 29.60 | 0.00 | 10→13 | 0.1466 | 32.07 | 0.00 |
| 2L-SiC | | | | 11→14 | 0.1491 | 30.03 | 0.00 |
| 7→9 | 0.1481 | 30.85 | 0.00 | 15 | 0.1554 | 0.00 | 9.15 |
| 10 | 0.1591 | 0.00 | 4.42 | 12→13 | 0.1395 | 4.70 | 0.00 |
| 8→9 | 0.1393 | 2.38 | 0.00 | 15 | 0.1506 | 27.68 | 0.00 |
| 10 | 0.1503 | 28.74 | 0.00 | | | | |

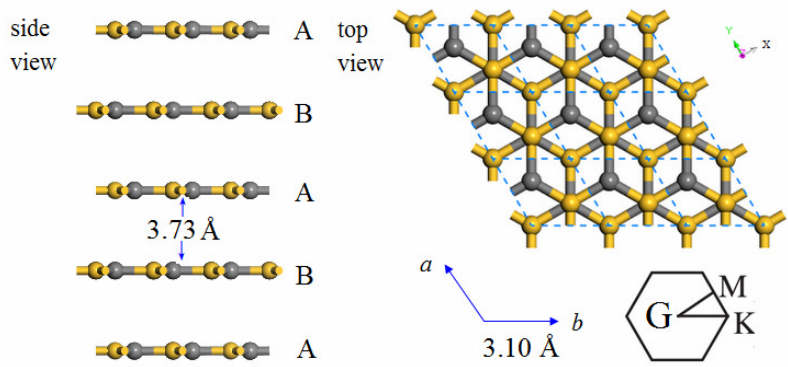

Figure 1. Crystal structures of 2D-SiC. The Cartesian coordinate system (top right) is for top viewed structure. The hexagonal lattice length ($a = b$) and interlayer distance are 3.10 and 3.73 Å, respectively. The unit cells are indicated by the dashed boxes.

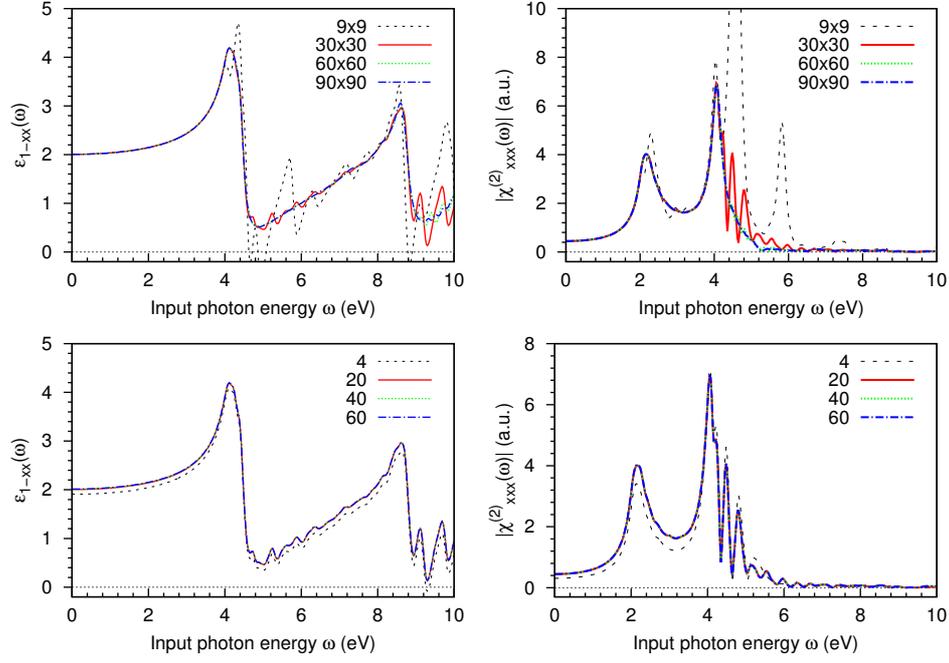

Figure 2. Convergence tests on the k-point mesh (up) with the fixed number of 20 empty states and on the number of empty states (down) with the fixed k-point mesh of 30×30×1 for $\varepsilon_1(\omega)$ and $|\chi^{(2)}_{xxx}(\omega)|$ of ML-SiC.

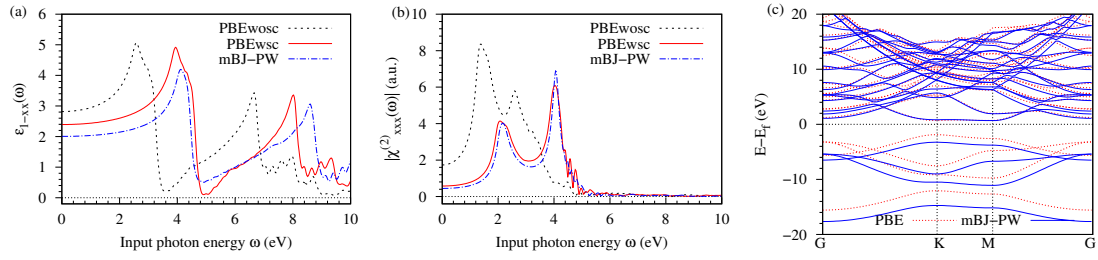

Figure 3. (a) $\varepsilon_{1\text{-}xx}(\omega)$ and (b) $|\chi^{(2)}_{xxx}(\omega)|$ of ML-SiC based on the PBE-wosc, PBE-wsc, and mBJ-PW with the all-electron FP-LAPW calculations. (c) Band structures of ML-SiC based on the PBE and mBJ-PW with the all-electron FP-LAPW calculations. The Fermi energy was set to zero.

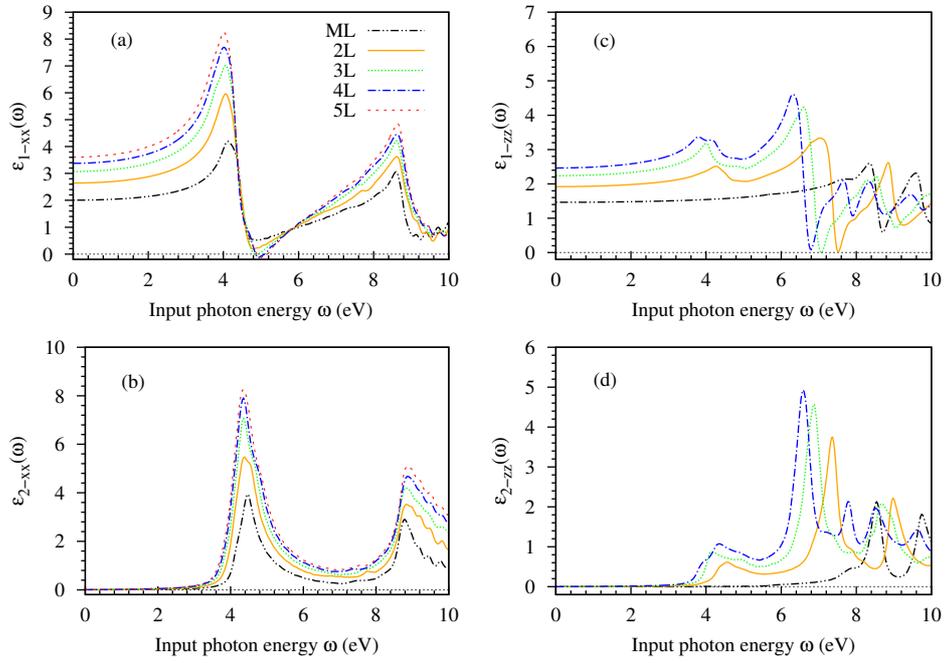

Figure 4. Real and imaginary parts of the dielectric constants (Eq.3) for ML-SiC and MuL-SiC.

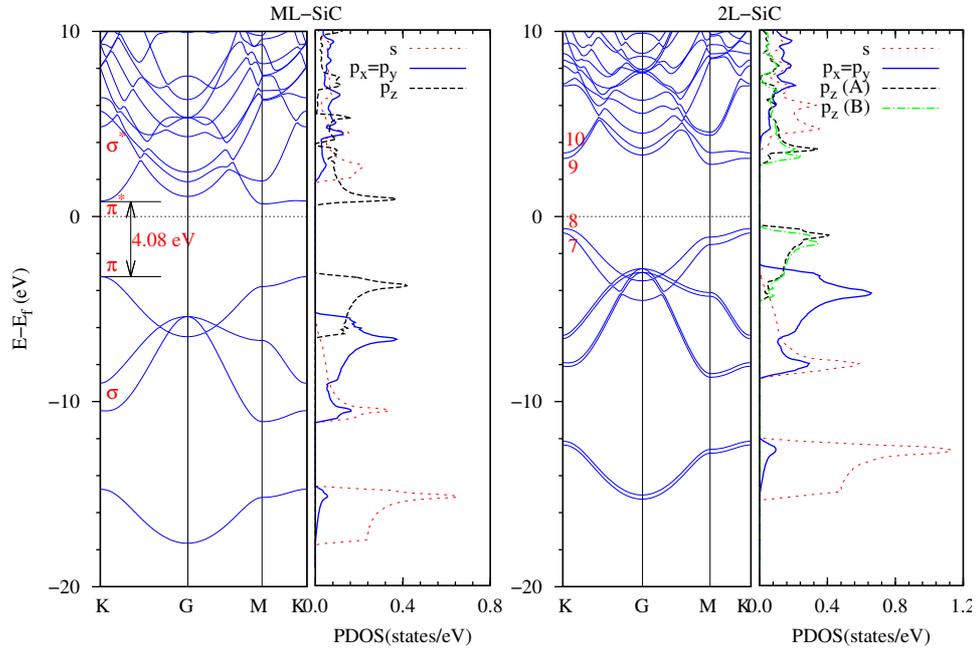

Figure 5. Band structures and partial density of states (PDOS) of ML-SiC and 2L-SiC. The "7, 8, 9, and 10" indicate the 7th, 8th, 9th, and 10th state, respectively. The "$p_z$ (A)" means that $p_z$ comes from the atoms of layer A (Fig.1).

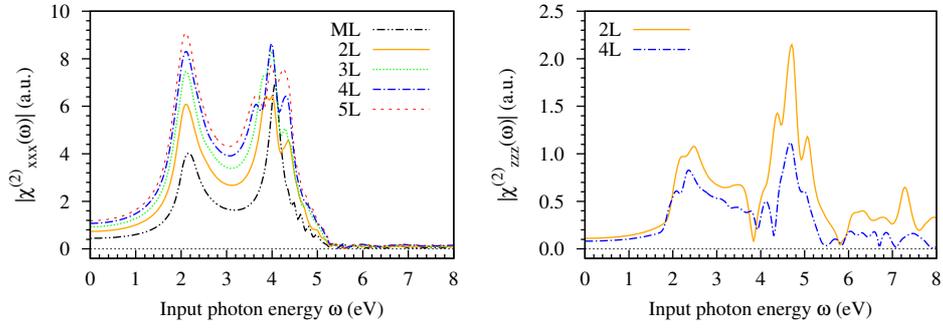

Figure 6. $|\chi^{(2)}(\omega)|$ along two polarization directions (*x* and *z*) for ML-SiC and MuL-SiC. Note that ML-SiC, 3L-SiC, and 5L-SiC have no nonzero $\chi^{(2)}_{zzz}(\omega)$ owing to the limitation of $D_{3h}$ symmetry.

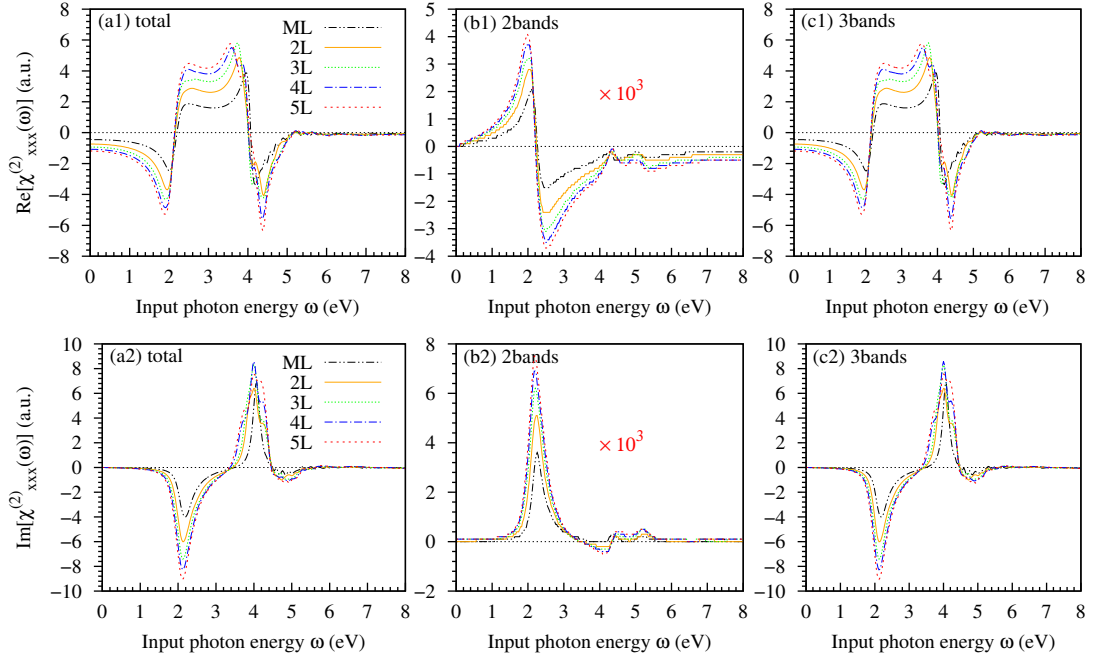

Figure 7. Real and imaginary parts of (a1 and a2) $\chi^{(2)}_{xxx}(\omega)$ (total) and (b1, b2, c1, and c2) $\chi^{(2)}_{xxx}(\omega)$ coming from two- (2bands) and three-band (3bands) terms (total = 2bands + 3bands). The two-band contribution is magnified by $\times 10^3$ for convenience of comparison.

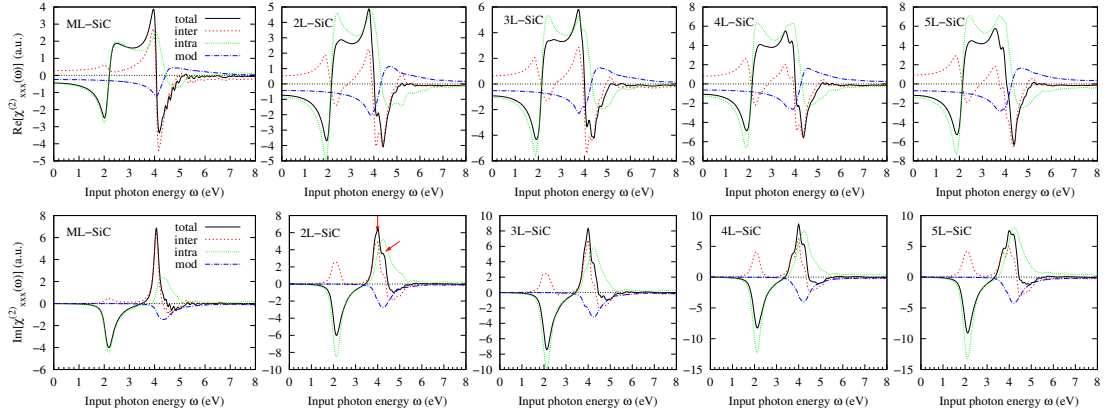

Figure 8. Real and imaginary parts of $\chi^{(2)}_{total}(\omega)$, $\chi^{(2)}_{inter}$, $\chi^{(2)}_{intra}$, and $\chi^{(2)}_{mod}$ for $\chi^{(2)}_{xxx}(\omega)$. Note that $\chi^{(2)}_{total}(\omega) = \chi^{(2)}_{inter} + \chi^{(2)}_{intra} + \chi^{(2)}_{mod}$.

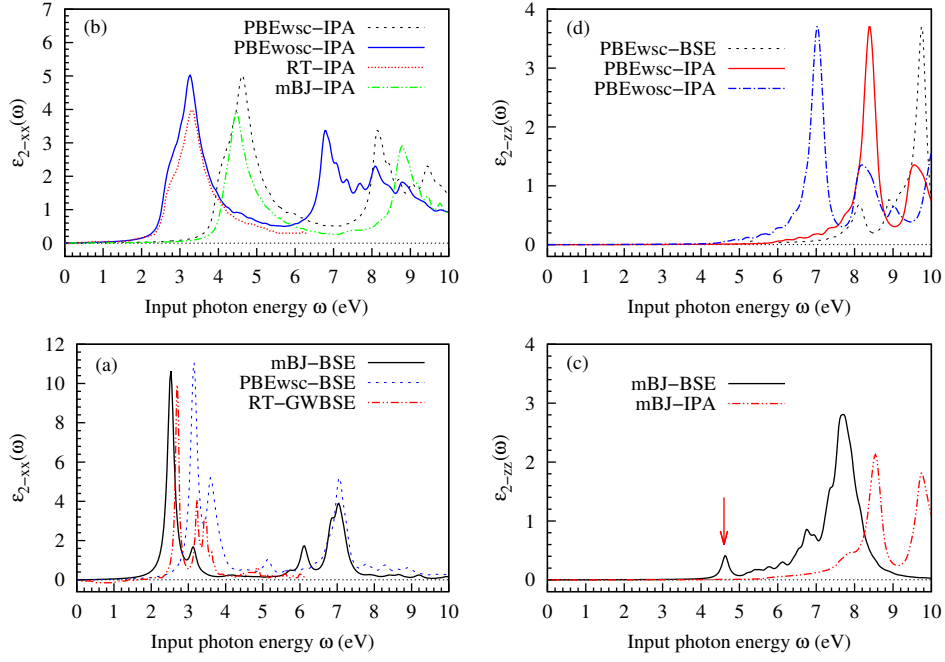

Figure 9. $\varepsilon_2(\omega)$ of ML-SiC based on different theoretical calculations including the real time (RT) calculations [7] within the IPA and GW+Bethe-Salpeter equation (GWBSE) framework, both mBJ and PBEwsc within the BSE framework, as well as mBJ, PBEwsc, and PBEwosc within the IPA.

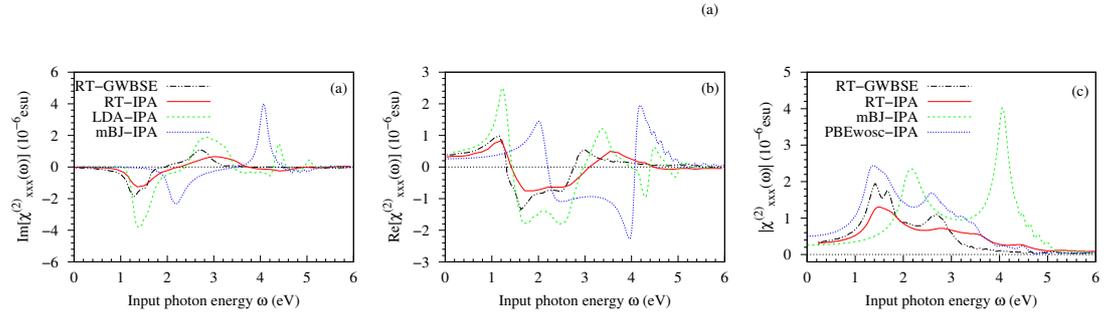

Figure 10. $\chi^{(2)}_{xxx}(\omega)$ of ML-SiC based on different theoretical calculations including the real time (RT) calculations [7] within the IPA and GW+Bethe-Salpeter equation (GWBSE) framework, as well as mBJ and PBEwosc within the IPA.